# Multiparty $d$-dimensional quantum information splitting

Andrzej Grudka* and Antoni Wójcik**

Faculty of Physics, Adam Mickiewicz University,

Umultowska 85, 61-614 Poznań, Poland

Abstract

Generalization of quantum information splitting protocol from qubits to qudits (quantum $d$-dimensional systems) is presented.

Recently Karlsson and Bourennane [1] and Hillery, Buzek and Berthiaume [2] have presented a scheme for quantum information splitting, which allows encoding of a state of a given particle (qubit) into $N$-particle state in such a way that all $N$ particles are necessary for perfect reconstruction of the original state. In their scheme it is the phase information which is absent in the state of any single particle, and moreover, in the state of any subset of $M$ particles except $M = N$. On the other, in the last few years there has been growing interest in generalization of quantum information protocols from two-dimensional systems (qubits) to $d$-dimensional systems (qudits). For example generalization of teleportation was presented in [3, 4, 5, 6], of quantum state purification in [7], of quantum cloning in [8]. Generalization of quantum key distribution protocols was proposed in [9, 10] and it was proven that the security of quantum cryptography increases with the dimension of the system used [11, 12]. It is the aim of our paper to present the $d$-dimensional generalization of the quantum secret sharing protocol.

Let us first briefly outline the scheme of Karlsson and Bourennane [1]. Alice is supposed to posses one qubit in a pure state $a|0\rangle + b|1\rangle$. She wants to teleport it to two parties $Bob_1$ and $Bob_2$ in such a way that none of them can reconstruct the original state by himself. However, when they cooperate the reconstruction is possible even if we allow only local operations and classical communication between them. This task can be easily performed when all three parties have the GHZ-state, i.e. maximally entangled three-particle state of the form

$$|SO(3)\rangle = \frac{1}{\sqrt{2}}\left(|0\rangle_1|0\rangle_2|0\rangle_3 + |1\rangle_1|1\rangle_2|1\rangle_3\right), \quad (1)$$

where particle 3 belongs to Alice and particles 1 and 2 belong to $Bob_1$ and $Bob_2$, respectively. In the first step of this protocol Alice performs the measurement of her particles (the qubit to be teleported and the qubit from GHZ-state) in the Bell basis. Depending on the result of measurement, the two remaining particles are in one of the four states

$$a|0\rangle_1|0\rangle_2 + b|1\rangle_1|1\rangle_2, \quad (2)$$

$$a|1\rangle_1|1\rangle_2 + b|0\rangle_1|0\rangle_2, \quad (3)$$

$$a|0\rangle_1|0\rangle_1 - b|1\rangle_1|1\rangle_2, \quad (4)$$

$$a|1\rangle_1|1\rangle_2 - b|0\rangle_1|0\rangle_2. \quad (5)$$

The last three states can be transformed to the first one by the use of local unitary operations, so we consider only this state. To recover the original state, Bob$_2$ applies Hadamard transform to his qubit and makes a measurement. The resulting state after the measurement is

$$(a|0\rangle_1 + b|1\rangle_1)|0\rangle_2 \tag{6}$$

or

$$(a|0\rangle_1 - b|1\rangle_1)|1\rangle_2. \tag{7}$$

Thus, if Bob$_2$ sends the result of his measurement to Bob$_1$ then Bob$_1$ can recover the original state.

If we replace the three-particle maximally entangled state in the protocol of teleportation by $N+1$-particle maximally entangled, then the above protocol can be easily generalized to split the information among $N$ parties (Bob$_1$, Bob$_2$,...,Bob$_N$). At the end of the splitting protocol $N$ parties share the state given by

$$a|0\rangle_1|0\rangle_2...|0\rangle_N + b|1\rangle_1|1\rangle_2...|1\rangle_N. \tag{8}$$

To recover the information Bob$_2$, Bob$_3$,...,Bob$_N$ perform Hadamard transformation and a measurement on their qubits. Knowing the results of all measurements, Bob$_1$ can transform his qubit to the state

$$a|0\rangle_1 + b|1\rangle_1. \tag{9}$$

Now we will generalize this protocol for qudits. We will need the following unitary operations. The first one is quantum Fourier transform $QFT_\mu$ which acts only on $\mu$-th qudit

$$QFT_\mu |j\rangle_\mu = \frac{1}{\sqrt{d}} \sum_{k=0}^{d-1} \exp\left(\frac{i2\pi jk}{d}\right) |k\rangle_\mu. \tag{10}$$

For the case of qubits ($d=2$) this is simply the Hadamard transform. The next one is a $d$-dimensional generalization of $XOR$ (performed on the $\mu$-th and $\nu$-th qudit), given by the following definition

$$XOR_{\mu\nu}|j\rangle_\mu|k\rangle_\nu = |j\rangle_\mu|k+j \bmod d\rangle_\nu. \tag{11}$$

Let $N+1$ parties share initially the state of the form

$$|SO(N+1)\rangle = \frac{1}{\sqrt{d}}\sum_{j=0}^{d-1}|j\rangle_1|j\rangle_2\ldots|j\rangle_{N+1}, \tag{12}$$

which is maximally entangled state of $N+1$ qudits. Alice also has a qudit in the state

$$|\Psi\rangle_0 = \sum_{j=0}^{d-1}\alpha_j|j\rangle_0. \tag{13}$$

Thus, the state of the whole system is

$$|\Psi\rangle_0|SO(N+1)\rangle = \frac{1}{\sqrt{d}}\sum_{k=0}^{d-1}\sum_{j=0}^{d-1}\alpha_k|k\rangle_0|j\rangle_1|j\rangle_2\ldots|j\rangle_{N+1}. \tag{14}$$

In order to teleport the state $|\Psi\rangle$, Alice and Bob$_1$, Bob$_2$,...,Bob$_N$ have to perform the following operations

(Step 1.) Alice applies $XOR_{0\,N+1}$ and then $QFT_0$. The resulting state is

$$\frac{1}{d}\sum_{k=0}^{d-1}\sum_{j=0}^{d-1}\sum_{l=0}^{d-1}\alpha_k \exp\left(\frac{i2\pi kl}{d}\right)|l\rangle_0|j+k \bmod d\rangle_1|j\rangle_2\ldots|j\rangle_{N+1}. \tag{15}$$

With the use of the substitution $k+j \bmod d = m$

Eq. 15 takes the form

$$\frac{1}{d}\sum_{k=0}^{d-1}\sum_{m=0}^{d-1}\sum_{l=0}^{d-1}\alpha_k \exp\left(\frac{i2\pi kl}{d}\right)|l\rangle_0|m-k \bmod d\rangle_1|m-k \bmod d\rangle_2\ldots|m-k \bmod d\rangle_N|m\rangle_{N+1}. \tag{16}$$

(Step 2.) Alice measures her qudits (in computational basis). If the result of this measurement is $|l\rangle_0|m\rangle_{N+1}$ then the remaining $N$ qudits are in the state

$$\sum_{k=0}^{d-1}\alpha_k \exp\left(\frac{i2\pi kl}{d}\right)|m-k \bmod d\rangle_1|m-k \bmod d\rangle_2\ldots|m-k \bmod d\rangle_N. \tag{17}$$

Steps 1 and 2 are equivalent to the measurement in the generalized Bell basis.

(Step 3.) Alice sends the results of her measurement $l$ and $m$ to Bob$_1$ and $m$ to Bob$_2$, Bob$_3$,...,Bob$_N$.

(Step 4.) Bob$_1$ performs unitary operation

$$|m - k \bmod d\rangle_1 \to \exp\left(\frac{-i2\pi kl}{d}\right)|k\rangle_1. \qquad (18)$$

Bob$_2$, Bob$_3$,...,Bob$_N$ perform unitary operations

$$|m - k \bmod d\rangle_\mu \to |k\rangle_\mu \quad (\mu = 2,...,N). \qquad (19)$$

The final state shared by Bob$_1$, Bob$_2$,...,Bob$_N$ is

$$|QSS(N)\rangle = \sum_{k=0}^{d-1} \alpha_k |k\rangle_1 ... |k\rangle_N. \qquad (20)$$

Each party now possesses a qudit, which can be described by the following reduced diagonal density operator

$$\rho_R = \sum_{k=0}^{d-1} |\alpha_k|^2 |k\rangle\langle k|, \qquad (21)$$

which means that each party alone does not have the information on phases between the states $|k\rangle$.

One of the parties (e.g. Bob$_1$) can reconstruct the original state if all parties cooperate in the process of the subsequent transforms

$$|QSS(N)\rangle \to |QSS(N-1)\rangle \to ... \to |QSS(1)\rangle = |\Psi\rangle_1. \qquad (22)$$

Below, we present how the transform $|QSS(K)\rangle \to |QSS(K-1)\rangle$ ($K = 2,...,N$) can be performed with the use of local operations and classical communication.

(Step 1.) Bob$_K$ applies quantum Fourier transform (QFT$_K$) to his qudit

$$QFT_K |QSS(K)\rangle = \frac{1}{\sqrt{d}} \sum_{j=0}^{d-1} \sum_{k=0}^{d-1} \alpha_j \exp\left(\frac{i2\pi jk}{d}\right) |j\rangle_1 |j\rangle_2 ... |k\rangle_K. \qquad (23)$$

(Step 2.) Bob$_K$ measures his qudit in the computational basis. If he obtains the result $k_K$ the state of the remaining qudits is

$$\overline{|QSS(K-1)\rangle} = \sum_{j=0}^{d-1} \alpha_j \exp\left(\frac{i2\pi jk_K}{d}\right) |j\rangle_1 |j\rangle_2 ... |j\rangle_{N-1}. \qquad (24)$$

(Step 3.) Bob$_K$ sends the result of his measurement to Bob$_1$.

(Step 4.) Bob$_1$ changes phase of his qudit in the following way

$$|j\rangle_1 \to \exp\left(\frac{-i2\pi jk_K}{d}\right)|j\rangle_1. \qquad (25)$$

After these steps the first $K-1$ parties share the following state

$$|QSS(K-1)\rangle = \sum_{j=0}^{d-1} \alpha_j |j\rangle_1 |j\rangle_2 \cdots |j\rangle_{K-1}. \qquad (26)$$

It is clear that if Bob$_2$, Bob$_3$,...,Bob$_N$ perform the above protocol (Step 1. To Step. 4) then Bob$_1$ will obtain the desired state of his qudit i.e. $|\Psi\rangle_1 = |QSS(1)\rangle = \sum_{j=0}^{d-1} \alpha_j |j\rangle_1$. It is worth noting that the transform $|QSS(N)\rangle \to |\Psi\rangle_1$ although written as a recursive formula (Eq. 22) can be performed in parallel. In this case, Bob$_1$ can perform all necessary phase changes in one step (instead of $N-1$). The appropriate unitary transform is

$$|j\rangle_1 \to \exp\left(\frac{-i2\pi j(k_2+k_3+...+k_N)}{d}\right)|j\rangle_1. \qquad (27)$$

In conclusion, we have presented a protocol allowing to split the information on the quantum state of a $d$ - dimensional particle among $N$ parties.

*Email address: agie@amu.edu.pl

**Email address: antwoj@amu.edu.pl